\input{aipcheck}

\documentclass[
    ,final            % use final for the camera ready runs
  ]
  {aipproc}

\layoutstyle{8x11single}

\def\BBz{0$\nu\beta\beta$}

\def\mj{M{\sc ajo\-ra\-na}}
\def\dem{D{\sc e\-mon\-strat\-or}}

\def\QBB{Q$_{\beta\beta}$}

\begin{document}

\title{Status of the \textsc{Majorana Demonstrator}}

% insert suggested PACS numbers in braces on next line
% 23.40.Hc Relation with nuclear matrix elements and nuclear structure
% 23.40-s beta decay, double beta decay, electron and muon capture
%23.40.Bw Weak-interaction and lepton (including neutrino) aspects
%21.10.Tg Lifetimes, widths
%27.50.+e 59 ² A ² 89
%14.60.Pq Neutrino mass and mixing
%\pacs{23.40-s, 23.40.Bw, 14.60.Pq, 27.50.+j}

\classification{23.40-s, 23.40.Bw, 14.60.Pq, 27.50.+j}
\keywords      {neutrinoless double-beta decay, germanium detector, Majorana neutrino mass}

\newcommand{\blhill}{Department of Physics, Black Hills State University, Spearfish, SD, USA}
\newcommand{\ITEP}{Institute for Theoretical and Experimental Physics, Moscow, Russia}
\newcommand{\JINR}{Joint Institute for Nuclear Research, Dubna, Russia}
\newcommand{\lbnl}{Nuclear Science Division, Lawrence Berkeley National Laboratory, Berkeley, CA, USA}
\newcommand{\lanl}{Los Alamos National Laboratory, Los Alamos, NM, USA}
\newcommand{\uw}{Center for Experimental Nuclear Physics and Astrophysics, and Department of Physics, University of Washington, Seattle, WA, USA}
\newcommand{\unc}{Department of Physics and Astronomy, University of North Carolina, Chapel Hill, NC, USA}
\newcommand{\duke}{Department of Physics, Duke University, Durham, NC, USA}
\newcommand{\ornl}{Oak Ridge National Laboratory, Oak Ridge, TN, USA}
\newcommand{\ou}{Research Center for Nuclear Physics and Department of Physics, Osaka University, Ibaraki, Osaka, Japan}
\newcommand{\pnnl}{Pacific Northwest National Laboratory, Richland, WA, USA}
\newcommand{\ttu}{Tennessee Tech University, Cookeville, TN, USA}
\newcommand{\sdsmt}{South Dakota School of Mines and Technology, Rapid City, SD, USA}
\newcommand{\usc}{Department of Physics and Astronomy, University of South Carolina, Columbia, SC, USA}
\newcommand{\usd}{Department of Physics, University of South Dakota, Vermillion, SD, USA}
\newcommand{\ut}{Department of Physics and Astronomy, University of Tennessee, Knoxville, TN, USA}
\newcommand{\tunl}{Triangle Universities Nuclear Laboratory, Durham, NC, USA}

\author{C. Cuesta$^{a}$, N.~Abgrall$^{b}$, I.J.~Arnquist$^{c}$, F.T.~Avignone~III$^{d,e}$, C.X.~Baldenegro-Barrera$^{e}$, A.S.~Barabash$^{f}$, F.E.~Bertrand$^{e}$, A.W.~Bradley$^{b}$, V.~Brudanin$^{g}$, M.~Busch$^{h,i}$, M.~Buuck$^{a}$, D.~Byram$^{j}$, A.S.~Caldwell$^{k}$, Y-D.~Chan$^{b}$, C.D.~Christofferson$^{k}$, P.-H.~Chu$^{o}$, J.A.~Detwiler$^{a}$, Yu.~Efremenko$^{l}$, H.~Ejiri$^{m}$, S.R.~Elliott$^{o}$, A.~Galindo-Uribarri$^{e}$, T.~Gilliss$^{n,i}$, G.K.~Giovanetti$^{n,i}$, J. Goett$^{o}$, M.P.~Green$^{e}$, J.~ Gruszko$^{a}$, I.S.~Guinn$^{a}$, V.E.~Guiseppe$^{d}$, R.~Henning$^{n,i}$, E.W.~Hoppe$^{c}$, S.~Howard$^{k}$, M.A.~Howe$^{n,i}$, B.R.~Jasinski$^{j}$, K.J.~Keeter$^{p}$, M.F.~Kidd$^{q}$, S.I.~Konovalov$^{f}$, R.T.~Kouzes$^{c}$, B.D.~LaFerriere$^{c}$, J.~Leon$^{a}$, J.~MacMullin$^{n,i}$, R.D.~Martin $^{j}$, R.~Massarczyk$^{o}$, S.J.~Meijer$^{n,i}$, S.~Mertens$^{b}$, J.L.~Orrell$^{c}$, C.~O'Shaughnessy$^{n,i}$, A.W.P.~Poon$^{b}$, D.C.~Radford$^{e}$, J.~Rager$^{n,i}$, K.~Rielage$^{o}$, R.G.H.~Robertson$^{a}$, E.~Romero-Romero$^{l,e}$, B.~Shanks$^{n,i}$, M.~Shirchenko$^{g}$, N.~Snyder$^{j}$, A.M.~Suriano$^{k}$, D.~Tedeschi$^{d}$, J.E.~Trimble$^{n,i}$, R.L.~Varner$^{e}$, S. Vasilyev$^{g}$, K.~Vetter$^{r,b}$, K.~Vorren$^{n,i}$, B.R.~White$^{e}$, J.F.~Wilkerson$^{n,i,e}$, C. Wiseman$^{d}$, W.~Xu$^{o}$, E.~Yakushev$^{g}$, C.-H.~Yu$^{e}$, V.~Yumatov$^{f}$, I.~Zhitnikov$^{g}$}
{address={$^{a}$ \uw   \\
          $^{b}$ \lbnl  \\
          $^{c}$  \pnnl \\
          $^{d}$  \usc \\
          $^{e}$  \ornl \\
          $^{f}$  \ITEP \\
          $^{g}$ \JINR  \\
          $^{h}$  \duke \\
          $^{i}$  \tunl\\
          $^{j}$  \usd \\
          $^{k}$  \sdsmt \\
          $^{l}$  \ut  \\
          $^{m}$  \ou  \\
          $^{n}$ \unc \\
          $^{o}$  \lanl \\
          $^{p}$  \blhill \\
          $^{q}$  \ttu \\
          $^{r}$  Alternate address: Department of Nuclear Engineering, University of California, Berkeley, CA, USA
          }}	

\begin{abstract}
The \mj\ Collaboration is constructing the \mj\ \dem, an ultra-low background, modular, HPGe detector array with a mass of 44-kg (29~kg $^{76}$Ge and 15 kg $^{nat}$Ge) to search for neutrinoless double beta decay in $^{76}$Ge.  The next generation of tonne-scale Ge-based neutrinoless double beta decay searches will probe the neutrino mass scale in the inverted-hierarchy region. The  \mj\ \dem\ is envisioned to demonstrate a path forward to achieve a background rate at or below 1 count/tonne/year in the 4~keV region of interest around the Q-value of 2039~keV. The~\mj~\dem~follows a modular implementation to be easily scalable to the next generation experiment. First, the prototype module was assembled; it has been continuously taking data from July 2014 to June 2015. Second, Module 1 with more than half of the total enriched detectors and some natural detectors has been assembled and it is being commissioned. Finally, the assembly of Module 2, which will complete \mj\ \dem, is already in progress.
\end{abstract}

\maketitle

%%%%%%%%%%%%%%%%%%%%%%%%%%%%%%%%%%%%%%%%%%%%
%% MAINMATTER
%%%%%%%%%%%%%%%%%%%%%%%%%%%%%%%%%%%%%%%%%%%%

\section{Introduction}
\label{sec1}

Neutrinoless double-beta (\BBz) decay is a model-independent method to search for lepton number violation and to determine the Dirac or Majorana nature of the neutrino~\cite{Zralek,Camilleri,Avignone,vergados}. Observation of this rare process would have significant implications for our understanding of the nature of neutrinos and matter in general. The 0$\nu\beta\beta$-decay rate for the light Majorana neutrino-mass process may be written as:

\begin{equation}\label{eq1}
\left(T^{0\nu}_{1/2}\right)^{-1}=G^{0\nu}|M_{0\nu}|^{2} \left( \frac{\langle m_{\beta\beta}\rangle}{m_{e}} \right)^{2}
\end{equation}
\noindent
where $G^{0\nu}$ is a phase space factor, $M_{0\nu}$ is a nuclear matrix element, $m_{e}$ is the electron mass,
and $\langle m_{\beta\beta}\rangle$ is the effective Majorana neutrino mass. The latter is given by

\begin{equation}\label{eq2}
\langle m_{\beta\beta}\rangle=\left|\sum_{i=0}^{3}U_{ei}^{2}m_{i}\right|
\end{equation}
\noindent
where $U_{ei}$ specifies the admixture of neutrino mass eigenstate $i$ in the electron neutrino. Then, assuming that 0$\nu\beta\beta$-decay is mainly driven by the exchange of light Majorana neutrinos, it is possible to establish an absolute scale for the neutrino mass, provided that the nuclear matrix elements are known.

Experimentally, \BBz-decay can be detected by searching the spectrum of the summed energy of the emitted betas for a monoenergetic line at the Q-value of the decay (\QBB). In previous-generation searches, the most sensitive limits on \BBz-decay came from the Heidelberg-Moscow experiment~\cite{Heil}, and the IGEX experiment~\cite{IGEX,IGEX2}, both using $^{76}$Ge. A direct observation of 0$\nu\beta\beta$-decay was claimed by a subgroup of the Heidelberg-Moscow collaboration~\cite{Kla}. Recent sensitive searches for~\BBz~have been carried out in $^{76}$Ge (GERDA~\cite{GERDA}) and $^{136}$Xe (KamLAND-Zen~\cite{KamLAND} and EXO-200~\cite{EXO,EXO2}), setting limits that do not support such a claim.

\section{Overview of the \mj~\dem}
\label{sec2}

The~\mj~\dem~\cite{mjd} is an array of enriched and natural germanium detectors that will search for the 0$\nu\beta\beta$-decay of $^{76}$Ge. The specific goals of the~\mj~\dem~are several: to demonstrate a path forward to achieving a background rate at or below 1~count/(ROI-t-y) in the 4~keV region of interest (ROI) around the 2039~keV~\QBB~of the $^{76}$Ge 0$\nu\beta\beta$-decay, when scaled up to a tonne scale experiment; show technical and engineering scalability toward a tonne-scale instrument; and perform searches for other physics beyond the Standard Model, such as dark matter and axions.

The experiment is composed of 44~kg of high-purity Ge (HPGe) detectors which also act as the source of $^{76}$Ge \BBz-decay. The benefits of HPGe detectors are that Ge is an intrinsically low-background source material, with understood enrichment chemistry, excellent energy resolution, and event reconstruction capabilities. P-type point contact detectors~\cite{ppc,ppc2} were chosen after extensive R\&D by the collaboration for their powerful background rejection capabilities. Twenty nine kg of the detectors are built from Ge material that is enriched to $>$87\% in $^{76}$Ge and 15~kg are fabricated from natural Ge (7.8\% $^{76}$Ge). The average mass of the enriched detectors is $\sim$850~g.

\begin {figure}[ht]
\includegraphics[width=0.65\textwidth]{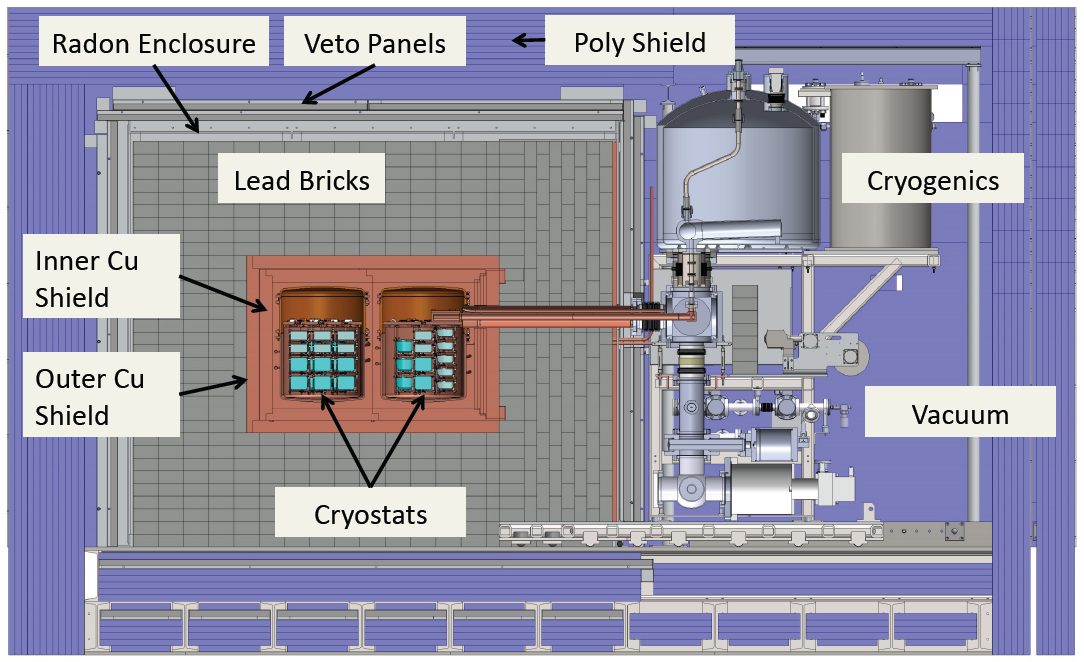}
\centering \caption{\it Schematic drawing of the \textsc{Majorana Demonstrator} shown with both modules installed.}
\label{fig:section}
\end {figure}

A modular instrument composed of two cryostats built from ultra-pure electroformed copper is being constructed. Each module hosts 7 strings of 3-5 detectors. The modules are operated in a passive shield that is surrounded by a 4$\pi$ active muon veto. To mitigate the effect of cosmic rays and prevent cosmogenic activation of detectors and materials, the experiment is being deployed at 4850~ft depth (4260~m.w.e. overburden) at the Sanford Underground Research Facility in Lead, SD~\cite{surf}. A schematic drawing of the \mj~\dem\ is shown in Figure~\ref{fig:section}.

The main technical challenge of the \mj\ \dem\ is to reach a background rate of 3~counts/(ROI-t-y) after analysis cuts, which projects to a background level of 1~count/(ROI-t-y) in a large scale experiment after accounting for additional improvements from thicker shielding, better self-shielding, and if necessary, increased depth. This background level represents a substantial improvement over previous generation experiments~\cite{GERDA}. To achieve this goal, background sources must be reduced and offline background rejection must be improved. The estimated ROI contributions based on achieved assays of materials sum to $<$3.5~counts/(ROI-t-y) in the \mj\ \dem\, and work is in progress to get the final estimate.

\section{The \mj~\dem\ implementation}
\label{sec3}

The~\mj~\dem~follows a modular implementation to be easily scalable to the next generation experiment. The modular approach allows the assembly and commissioning of each module independently, providing a fast deployment and minimum interference with already-operational detectors.

\begin{figure}
 \includegraphics[width=0.62\textwidth]{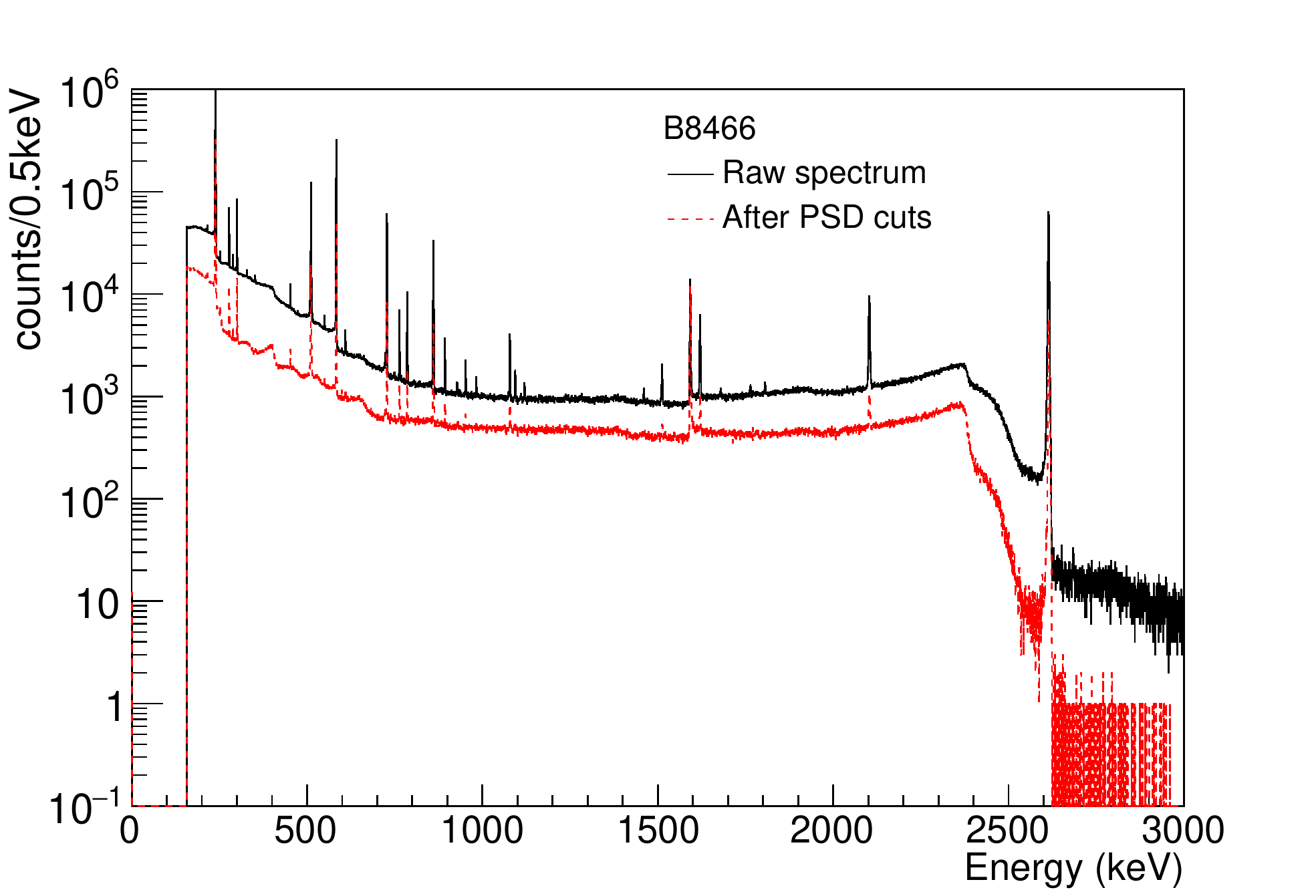}
  \caption{Energy spectra before and after the PSD cut on $^{228}$Th calibration data from detector B8466, which is a $^{nat}$Ge detector within the Prototype Module.}
  \label{fig:psd}
\end{figure}

As a first step, the prototype module, an initial prototype cryostat fabricated from commercially produced copper, was loaded with three strings of natural-abundance germanium detectors and placed into the shielding. It has taken data from June 2014 through July 2015. It has served as a test bench for mechanical designs, fabrication methods, and assembly procedures to be used for the construction of the electroformed-copper Modules~1~\&~2. The data analysis is in progress, but as an example of the results obtained, Figure~\ref{fig:psd} shows a $^{228}$Th calibration spectra before and after the pulse shape discrimination (PSD) cut in one of the prototype module detectors. The PSD cut selects single-site events with a $>$90\% efficiency using the study of the current peak amplitude to total energy~\cite{AEgerda}. Multiple interaction site events from gamma rays (non-plausible \BBz-decay candidate events) are rejected with a $>$90\% efficiency. Then, the PSD cut reduces the continuum background during a $^{228}$Th calibration at ROI $>$50\%, as shown in Figure~\ref{fig:psd}.

The second step of deployment is Module~1 which has already been assembled. Module~1 houses 17~kg of enriched germanium detectors and 6~kg of natural germanium detectors. The strings were assembled and characterized in dedicated String Test Cryostats. Then, Module~1 was moved into the shield; it is taking commissioning data at the moment. Figure~\ref{fig:section} shows Module~1 detectors and Module~1 being moved to the shield. Module~1 data taking, searching for \BBz-decay and including data blindness is foreseen to start soon. Finally, the last step is Module~2 which is composed of 12~kg of enriched and 9~kg of natural Ge detectors. The assembly of Module~2 has already started and its commissioning will take place by the end of 2015.

\begin {figure}[ht]
\includegraphics[height=0.27\textheight]{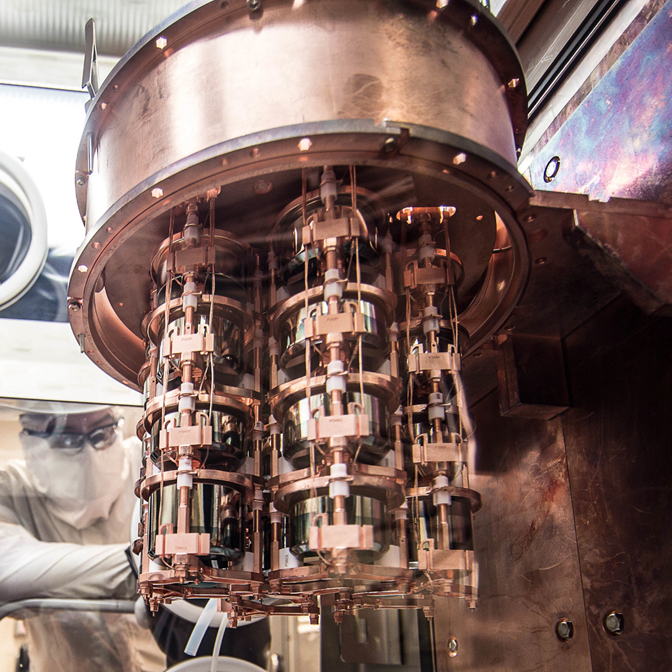}
\includegraphics[height=0.27\textheight]{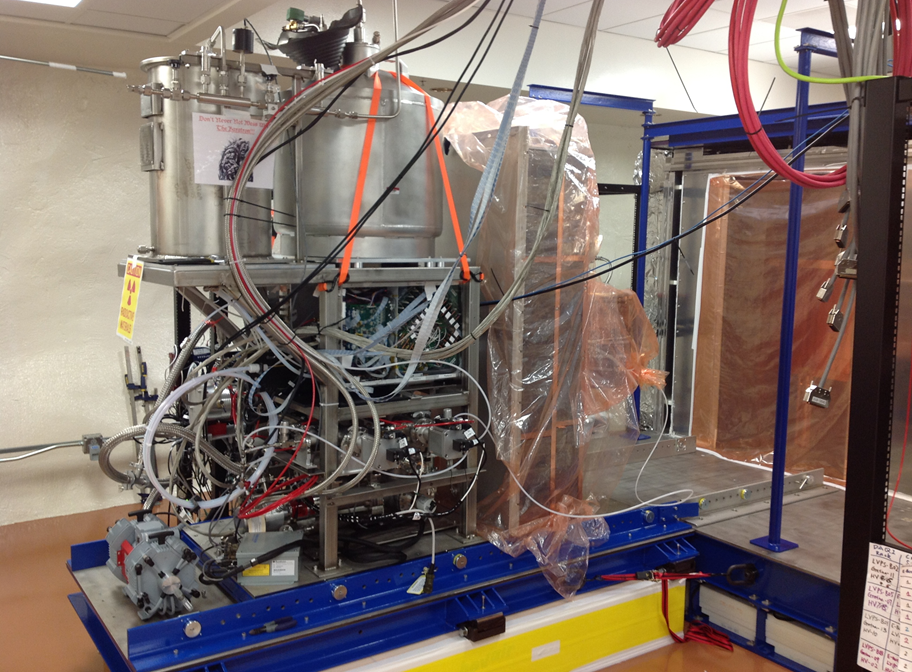}
\centering \caption{\it Module~1 detectors (left) and Module~1 being moved into the shield (right).}
\label{fig:string}
\end {figure}

%%%%%%%%%%%%%%%%%%%%%%%%%%%%%%%%%%%%%%%%%%%%%%%%
%% BACKMATTER
%%%%%%%%%%%%%%%%%%%%%%%%%%%%%%%%%%%%%%%%%%%%%%%%

\begin{theacknowledgments}

This material is based upon work supported by the U.S. Department of Energy, Office of Science, Office of Nuclear Physics. We acknowledge support from the Particle Astrophysics Program of the National Science Foundation. This research uses these US DOE Office of Science User Facilities: the National Energy Research Scientific Computing Center and the Oak Ridge Leadership Computing Facility. We acknowledge support from the Russian Foundation for Basic Research. We thank our hosts and colleagues at the Sanford Underground Research Facility for their support.

\end{theacknowledgments}

%%%%%%%%%%%%%%%%%%%%%%%%%%%%%%%%%%%%%%%%%%%%%%%%
%% The bibliography can be prepared using the BibTeX program or
%% manually.
%%
%% The code below assumes that BibTeX is used.  If the bibliography is
%% produced without BibTeX comment out the following lines and see the
%% aipguide.pdf for further information.
%%
%% For your convenience a manually coded example is appended
%% after the \end{document}
%%%%%%%%%%%%%%%%%%%%%%%%%%%%%%%%%%%%%%%%%%%%%%%%

%%%%%%%%%%%%%%%%%%%%%%%%%%%%%%%%%%%%%%%%%%%%%%%%
%% You may have to change the BibTeX style below, depending on your
%% setup or preferences.
%%
%%
%% For The AIP proceedings layouts use either
%%%%%%%%%%%%%%%%%%%%%%%%%%%%%%%%%%%%%%%%%%%%

\bibliographystyle{aipproc}   % if natbib is available
%\bibliographystyle{aipprocl} % if natbib is missing

%%%%%%%%%%%%%%%%%%%%%%%%%%%%%%%%%%%%%%%%%%%
%% You probably want to use your own bibtex database here
%%%%%%%%%%%%%%%%%%%%%%%%%%%%%%%%%%%%%%%%%%%
\bibliography{CCuesta_MEDEX15}

%%%%%%%%%%%%%%%%%%%%%%%%%%%%%%%%%%%%%%%%%%%
%% Just a reminder that you may have to run bibtex
%% All of it up to \end{document} can be removed
%% if you don't like the warning.
%%%%%%%%%%%%%%%%%%%%%%%%%%%%%%%%%%%%%%%%%%%
\IfFileExists{\jobname.bbl}{}
 {\typeout{}
  \typeout{******************************************}
  \typeout{** Please run "bibtex \jobname" to optain}
  \typeout{** the bibliography and then re-run LaTeX}
  \typeout{** twice to fix the references!}
  \typeout{******************************************}
  \typeout{}
 }

\end{document}